\documentclass[useAMS,usenatbib]{mn2e}

\usepackage{graphics,epsfig}
\usepackage{graphicx}
\usepackage{amssymb}
\usepackage{amsmath}

\title[GMRT observations of SBS 1129+576]{\ion{H}{i} studies of eXtremely Metal Deficient galaxies -- II. 
Giant Metrewave Radio Telescope observations of SBS~1129+576}
\author[Ekta, Jayaram N. Chengalur, S. A. Pustilnik]
{Ekta,$^1$\thanks{ekta@ncra.tifr.res.in,chengalu@ncra.tifr.res.in,sap@sao.ru}
Jayaram N. Chengalur,$^1$\footnotemark[1] 
Simon A. Pustilnik$^2$\footnotemark[1]\\
$^1$ National Centre for Radio Astrophysics, Post Bag 3, Ganeshkhind, Pune 411 007, India\\
$^2$ Special Astrophysical Observatory of RAS, Nizhnij Arkhyz, Karachai-Circasia 369167, Russia}

\newcommand{\kms}{km~s$^{-1}$}

\DeclareRobustCommand{\ion}[2]{%
\relax\ifmmode
\ifx\testbx\f@series
{\mathbf{#1\,\mathsc{#2}}}\else
{\mathrm{#1\,\mathsc{#2}}}\fi
\else\textup{#1\,{\mdseries\textsc{#2}}}%
\fi}

\begin{document}

\label{firstpage}

\date{Accepted 2006 July 31. Recevied 2006 July 20; in original form 2006 May 16}

\pagerange{\pageref{firstpage}--\pageref{lastpage}} \pubyear{2006}

\maketitle

\begin{abstract}

We present Giant Metrewave Radio Telescope \ion{H}{i} observations of an eXtremely Metal Deficient 
galaxy SBS~1129+576. SBS~1129+576 has a weighted mean oxygen 
abundance of 12~+~$\log$~(O/H)~=~7.41~$\pm$~0.07, or 1/18 of the solar value.
Our \ion{H}{i} observations show that the galaxy is strongly interacting
with a companion (projected separation $\sim$27~kpc) galaxy, 
SBS~1129+577. \ion{H}{i} emission from a third, smaller galaxy,
SDSS~J113227.68+572142.3, is also present in the data cube. 
We study the \ion{H}{i} morphology and kinematics of this small group
at angular resolutions ranging from $\sim$40 to 8~arcsec.
The low-resolution map shows a bridge of emission connecting
the two larger galaxies and a large one-armed spiral distortion of 
the disc of SBS~1129+577. We measure \ion{H}{i} masses of
$\sim$4.2~$\times$~10$^{8}$, $\sim$2.7$~\times$~10$^{9}$,
and $\sim$2.1~$\times$~10$^{8}$~M$_{\odot}$ for SBS~1129+576, SBS~1129+577
and the gas in the bridge, respectively. Assuming that most of the bridge
gas originally came from SBS~1129+576, approximately one-third of its original gas mass
has been stripped off. The third smaller galaxy has an \ion{H}{i} mass of
(M$_{\rm HI}$~$\sim$1.1$\times$10$^{7}$ M$_{\odot}$) and does not show
any sign of interaction with the other two galaxies.
The higher-resolution maps show that SBS~1129+577 has a central bar
and a ring surrounding the bar; there is also a hint of an integral-shaped 
warp in SBS~1129+576.  All
these features are very likely to have been induced by the 
tidal interaction. In both SBS~1129+576 and SBS~1129+577, there is, 
in general, a good correspondence between regions with high \ion{H}{i} 
column density and those with ongoing star formation. 
The two brightest \ion{H}{ii} regions in SBS~1129+576 have (inclination-corrected)
gas column densities of $\sim$1.6$\times$10$^{21}$ and  
$\sim$1.8$\times$10$^{21}$ atoms~cm$^{-2}$, respectively. The 
inclination-corrected \ion{H}{i} column density near the \ion{H}{ii} regions in
SBS~1129+577 is generally above $\sim$2.0$\times$10$^{21}$ atoms~cm$^{-2}$.
These values are close to the threshold density for star
formation observed in other blue compact galaxies.
In contrast to SBS~1129+576 and SBS~1129+577 which are very
gas-rich, the third member of this group, SDSS~J113227.68+572142.3, is
gas-poor.  
\end{abstract}


\section{INTRODUCTION}
\label{sec:intro}

  A very small fraction of low-mass, gas-rich galaxies have very low
observed interstellar medium (ISM) metallicities, 12~+~$\log$~(O/H)~$\leq$~7.65. About 50 such
galaxies are now known (see e.g. the review by \cite{KO2000}) and are
termed here as eXtremely Metal Deficient (XMD) galaxies. A few
XMD galaxies, for example, I~Zw~18 and SBS~ 0335$-$052 E and W, where there
is no evidence for an old stellar population, are considered as
good candidates for young galaxies in the local Universe.
In general, however, old stellar populations have been found in most 
(of the as of yet small numbers) of well-studied XMD galaxies. 
Despite the presence of an older stellar population, the very low 
ISM metallicities of XMDs indicate that they are relatively unevolved 
systems. Further evidence for their relatively unevolved status comes 
from the fact that they are very gas rich; neutral hydrogen is often 
the dominant baryonic component in these galaxies. The study of 
the neutral hydrogen content, distribution and kinematics is hence 
important in understanding this small, but interesting subclass 
of dwarf galaxies. As the best local analogues of young low-mass 
galaxies in the high-redshift Universe, detailed studies of the
connections between the gas distribution, kinematics and star
formation in these galaxies could provide useful input into
models of galaxy formation and evolution. A systematic, 
\ion{H}{i} survey of 22 XMD galaxies was conducted at the 
Nan\c {c}ay Radio Telescope (NRT) by \cite{MP05}. This survey
increased the number of XMD galaxies with available integrated 
\ion{H}{i} parameters by almost a factor of 2 (earlier surveys were
done by e.g. \cite{thuan99} 1999). Follow-up interferometric 
observations of a subsample of the galaxies observed at
the NRT are being done using the Giant Metrewave Radio 
Telescope (GMRT). SBS~1129+576, the subject of this paper, 
is the first object from this subsample.

   SBS~1129+576 is a very metal-poor (with weighted mean oxygen abundance 
12~+~$\log$~(O/H)~=~7.41~$\pm$~0.07; \cite{guseva03} 2003) nearby blue compact
galaxy (BCG). It was discovered as part of the Second Byurakan Survey (SBS)
(\cite{markarian83}; \cite{lipovetsky88}). \cite{Pustilnik01} measure 
$M_{\rm B} \sim$16.50 mag for this galaxy. \cite{guseva03} (2003) present $V$- and
$I$-band CCD images as well as spectroscopic data for this galaxy. The CCD
images show the galaxy to contain a chain of compact \ion{H}{ii} regions within an
elongated low surface brightness (LSB) envelope. The \ion{H}{ii} regions and the LSB
component both are bluer than typical for dwarf irregular or even blue
compact dwarf (BCD) galaxies. The blue colour and the very low metallicity
of SBS~1129+576 both are indicative of a relatively young age for the
dominant stellar population. However, as is generally the case with very
low metallicity galaxies, the spectroscopic and photometric data do not allow
one to rule out models in which the galaxy contains stellar populations as
old as 10~Gyr, albeit with the majority of the current luminosity
coming from a much younger ($\sim$100~Myr) population.

\cite{thuan99} (1999) presented NRT \ion{H}{i} data for SBS~1129+576. The measured 
heliocentric velocity is 1566~$\pm$~2 \kms, which corresponds to a
distance of 26.3 Mpc (after correction for Virgocentric infall).
\cite{thuan99} (1999) noted the possibility of confusion with a nearby 
dwarf irregular galaxy, SBS~1129+577 (MCG+10-17-010). \cite{pustilnik01} (2001a)
suggested that interaction with this companion probably triggered 
starburst in SBS~1129+576, whereas \cite{petrosian02} noted that SBS~1129+576 
has a peculiar disturbed morphology and classified it as a 'merger'. 

    We present here GMRT \ion{H}{i} 21-cm observations of emission from the
SBS~1129+576/577 pair. The rest of this paper is divided as follows. In
Section~\ref{sec:obs}, the observations and data reduction are described, while 
the results of the observations are described in Section~\ref{sec:res}. In
Section~\ref{sec:dis}, we compare SBS~1129+576 with the other very metal-poor
galaxies, and Section~\ref{sec:summary} provides a summary of our results.

\section{OBSERVATIONS AND DATA REDUCTION}
\label{sec:obs} 

   The GMRT observations, conducted on 2004 February 12 (see Table 1) had a total on-source
time of $\sim$7~h. The observations used a total bandwidth of 2.0 MHz
($\sim$425~\kms) centred at 1413.0~MHz. The FX correlator
provided 128 spectral channels, that is, a spectral resolution of $\sim$15.6~kHz
($\sim$3.3~\kms). Flux and bandpass calibration was done using short
observations of 3C~48 and 3C~286, while phase calibration was
done using either one of the VLA calibrator sources 1035+564 or 1219+584.
After flagging out bad-visibility points, flux and phase calibration was done
for a single spectral channel. This calibration was applied to the original
multichannel file, after which bandpass calibration was done. Finally, flux-,
phase- and bandpass-calibrated visibilities for the target source were split
out into a separate uv-file using the task {\small SPLIT}. Continuum emission was
subtracted from the visibilities using the task {\small UVSUB}. The GMRT has a hybrid 
configuration which allows images at a range of resolutions to be made 
from data from a single observing run. Spectral image cubes were hence made 
at several angular resolutions ranging from $\sim$40 to 
$\sim$8~arcsec. Integrated \ion{H}{i} emission as well as the 
\ion{H}{i} velocity field were derived using the task {\small MOMNT}. 

\begin{table}
\caption{Parameters of the observations using the GMRT}
\begin{tabular}{ll}
\hline
Date of observations & 2004 February 12\\
Field centre RA(J2000) & $11^{h}32^{m}02^{s}.507$ \\
Field centre Dec.(J2000) & 57\degr24\arcmin39\arcsec.960 \\
The velocity at the band centre & 1566~\kms \\
Time on source & 7~h \\
Number of channels & 128 \\
Channel separation & 3.3~\kms \\
Synthesized beams (FWHM in arcsec$^{2}$) & 42~$\times$~40, 
17~$\times$~14, 8~$\times$~7 \\
rms noise (mJy~beam$^{-1}$) & 1.6, 1.1, 1.0 \\
\hline
\end{tabular}
\end {table}

\section{RESULTS OF THE OBSERVATIONS}
\label{sec:res}

\subsection{Continuum emission}
\label{ssec:cont}

  No continuum radio emission was detected from the discs of either
SBS~1129+576 or SBS~1129+577. The brightest continuum source in the field is
87GB~112918.9+574015 (which lies $\sim$1.2~arcmin from the phase center)
for which we measure a flux density of 130~$\pm$~2~mJy. This source is 
coincident with the \ion{H}{i} bridge (see Section~\ref{ssec:line}), between the 
two galaxies. A search for \ion{H}{i} absorption against this source (in the 
high-resolution data cube) gives a 3$\sigma$ upper limit to the optical 
depth of 0.08 in the velocity range 1389--1706~\kms. If we take the 
\ion{H}{i} column density to be 4.6$\times$10$^{20}$~atoms~cm$^{-2}$(i.e. as obtained from 
the 40-arcsec-resolution \ion{H}{i} image at the position of the continuum 
source), this corresponds to a 3$\sigma$ lower limit of 956~K on 
the spin temperature.

\subsection{Line emission}
\label{ssec:line}

\begin{figure}
\begin{center}\includegraphics[width=7cm]{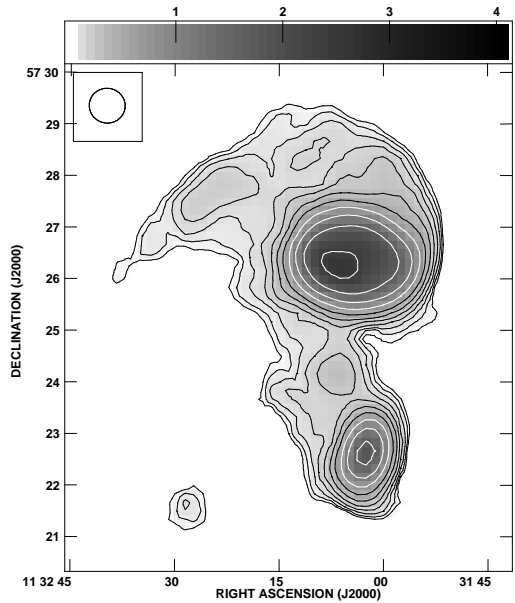}\end{center}
\caption{Integrated \ion{H}{i} emission map of the SBS~1129+576 field in contours. 
The resolution is 42~$\times$~40~arcsec$^{2}$ and the contours
are at 3.0, 4.5, 6.8, 10.2, 15.2, 22.8, 34.3, 51.4, 77.1, 115.6 
and 173.5~$\times$~10$^{19}$~atoms~cm$^{-2}$. The same map is shown in grey-scale 
over column density range of 2.6~$\times$~10$^{19}$--2.7~$\times$~10$^{21}$ 
atoms~cm$^{-2}$.}
\label{fig:mom0}
\end{figure}

The integrated \ion{H}{i} emission map of the SBS~1129+576 field at an angular
resolution of $\sim$42~$\times$~$\sim$40~arcsec$^{2}$ is shown in 
Fig.~\ref{fig:mom0}. There is a clear bridge of emission connecting the
two galaxies; SBS~1129+576 shows a hint of a tail to the south-west, while the
outer gas in SBS~1129+577 is distorted into a strong one-armed spiral pattern.
The total extent of \ion{H}{i} emission from the two galaxies is 
7.5~$\times$~5.6~kpc$^{2}$. \ion{H}{i} emission from a faint companion dwarf galaxy 
(SDSS~J113227.68+572142.3) at an angular distance of 3.5~arcmin to the south-east 
of SBS~1129+576 is also detected. 

  Channel maps at the same resolution are shown in Fig.~\ref{fig:chanA}. 
A counterarm can be seen to develop on the north-western
side of SBS~1129+577 at velocities $\sim$1562--1576~\kms. The north-eastern
part of this galaxy's disc also shows \ion{H}{i} extensions at a similar velocity
range; this may be related to the \ion{H}{i} in the bridge. At a velocity of
1592.7~\kms, these two arms are well developed and separated; however,
at still larger velocities they appear to merge to form the one-armed spiral
noted above. We detect \ion{H}{i} emission from SDSS~J113227.68+572142.3 in 
the velocity range 1526.0--1542.7~\kms. The emission is above the $\sim$3$\sigma$ level in 
all the channels in this range, except for the channel at $\sim$1529.3~\kms\ in 
which it is $\sim$2$\sigma$, and channel at 1539.3~\kms in which there is no emission 
at this position.

\begin{figure*}
\includegraphics[width=18.0cm]{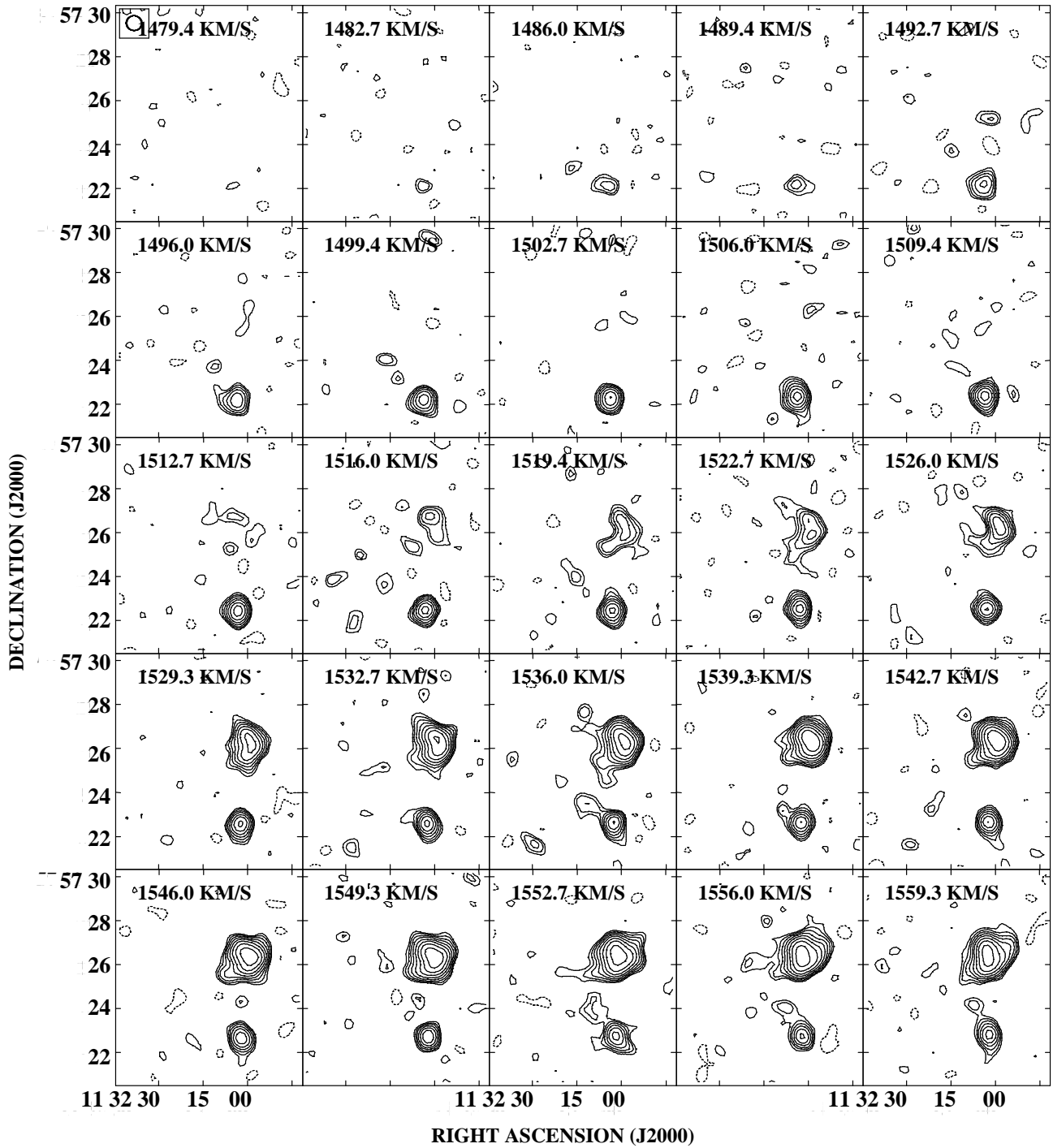}
\caption{Channel maps for SBS~1129+576 and SBS~1129+577 at the resolution
of 42~$\times$~40~arcsec$^{2}$. The contour levels are
at -3.2, 3.2, 4.5, 6.4, 9.1, 12.8, 18.1, 25.6, 36.2, 51.1 mJy~Bm$^{-1}$.}
\label{fig:chanA}
\end{figure*}
\setcounter{figure}{1}
\begin{figure*}
\includegraphics[width=18.0cm]{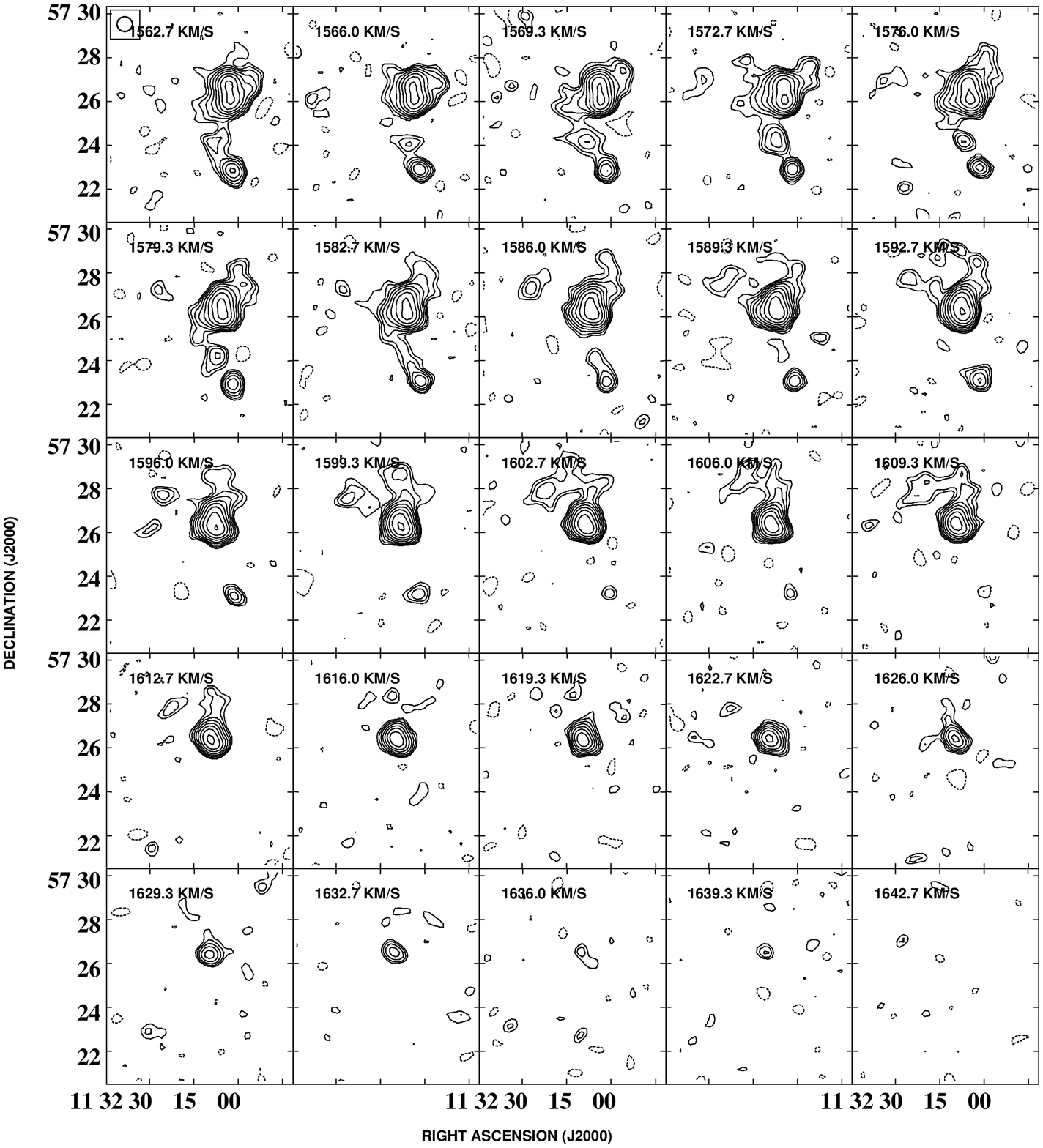}
\caption{{\it continued}}
\label{fig:chanB}
\end{figure*}

\begin{figure}
\includegraphics[width=6.75cm]{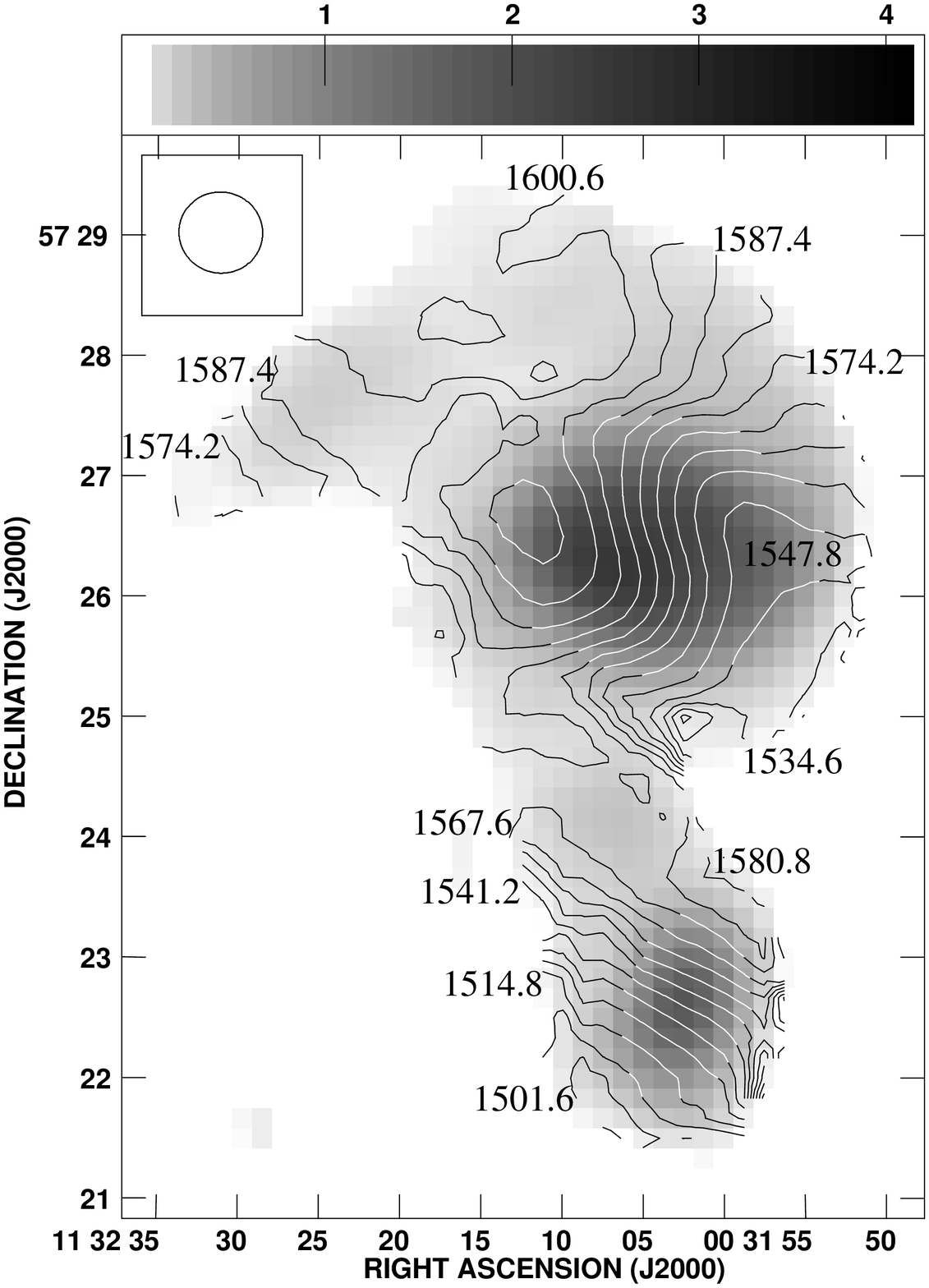}
\caption{Intensity-weighted velocity field, made from
42~$\times$~40-arcsec$^{2}$ resolution data. The velocity contours are
uniformly spaced with an interval of 6.6~\kms\ and range from 1495 to
1607~\kms. Integrated \ion{H}{i} emission map is shown in grey-scale over 
column density range of 2.6~$\times$~10$^{19}$--2.7~$\times$~10$^{21}$~atoms~cm$^{-2}$.}
\label{fig:mom1}
\end{figure}

\begin{figure}
\begin{center}\includegraphics[width=6.15cm,angle=270]{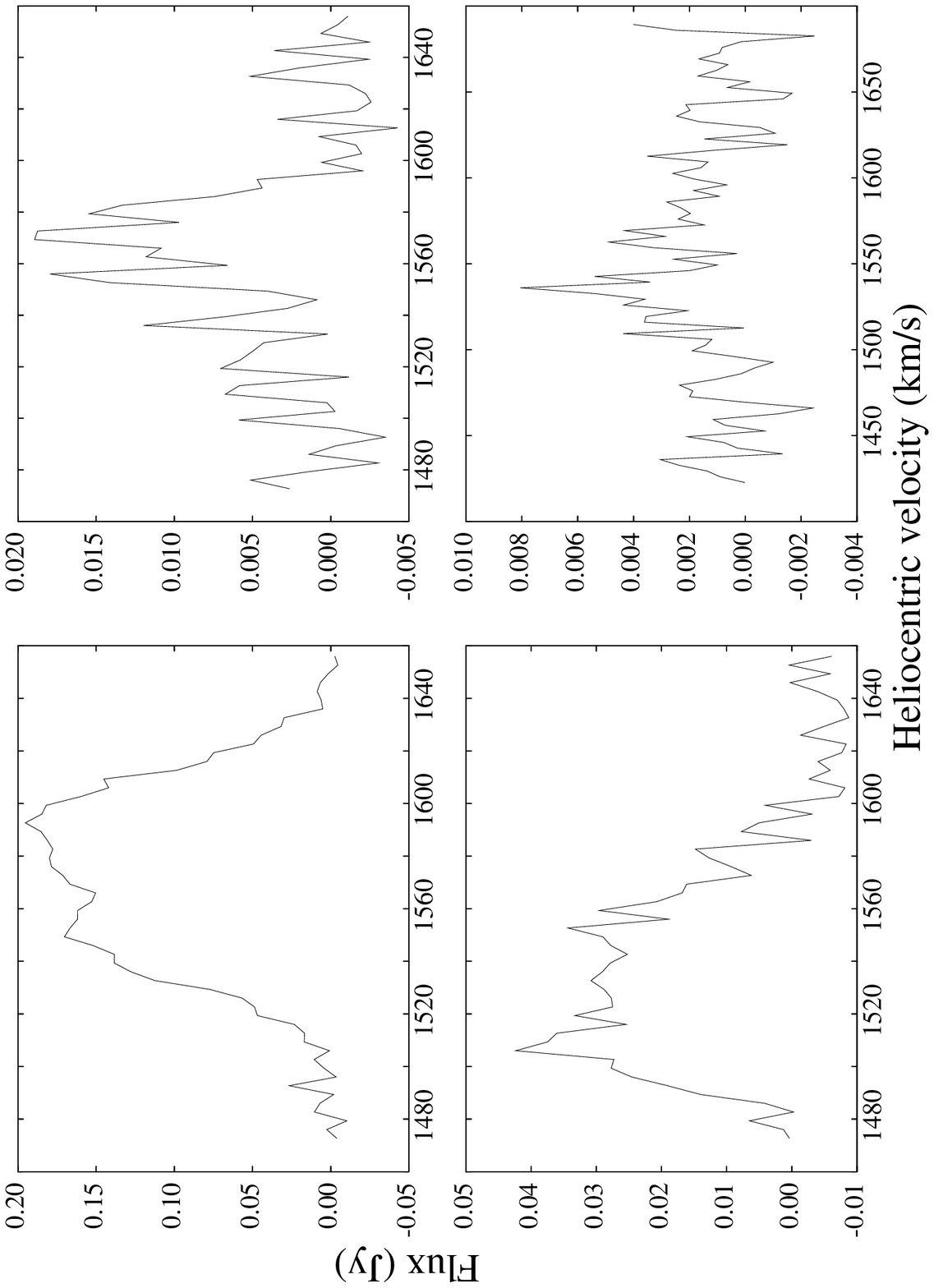}\end{center}
\caption{Integrated spectrum for the SBS~1129+577, bridge, SBS~1129+576 and
companion dwarf made from the 42~$\times$~40-arcsec$^{2}$-resolution data cube.}
\label{fig:spectra}
\end{figure}

\begin{figure}
\includegraphics[width=8.45cm]{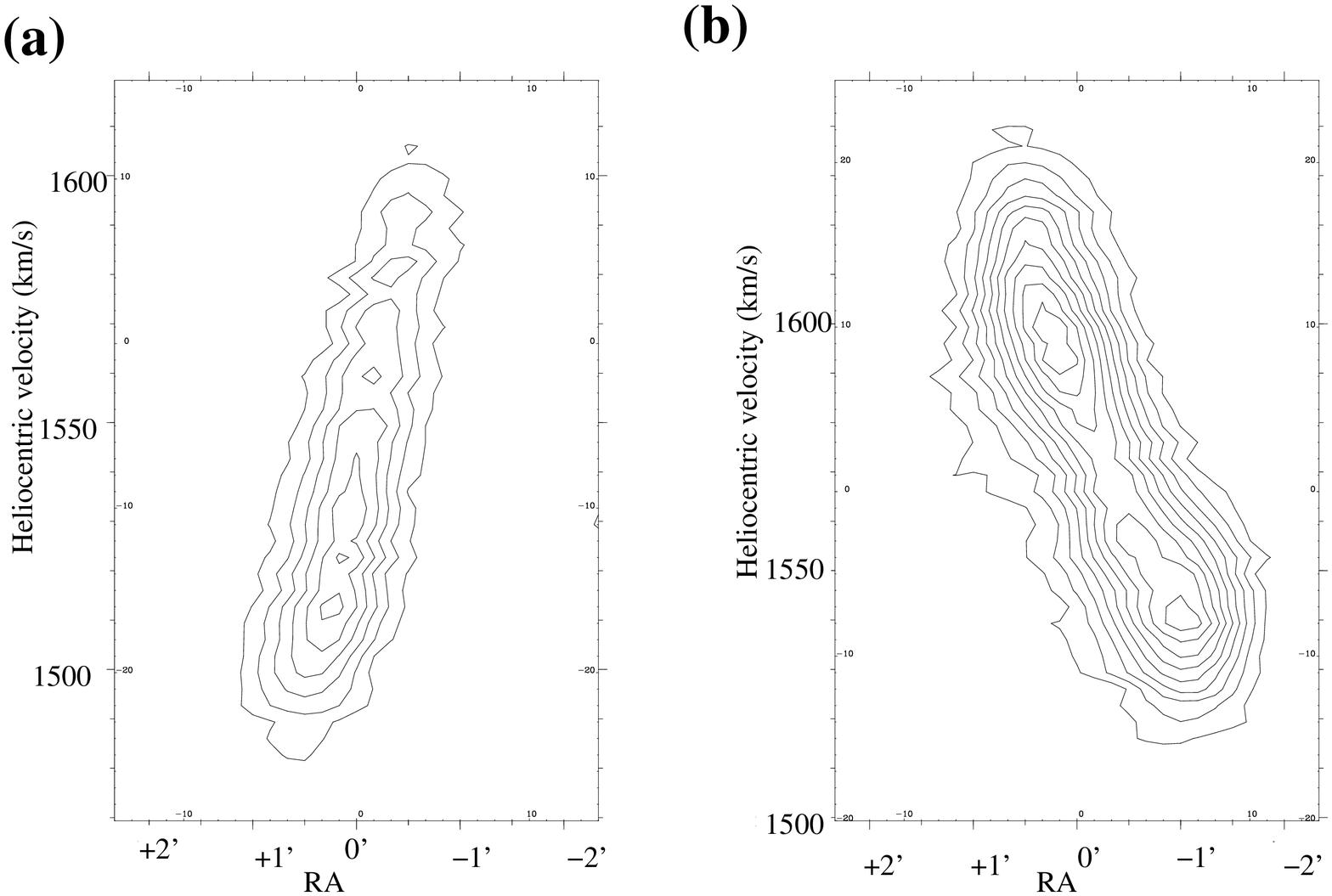}
\caption{\textbf{a)} Major axis P--V diagram for SBS~1129+576.
  The contour levels are at 5, 10, 15, 20, 25,
  29 mJy~beam$^{-1}$. \textbf{b)} Major axis P--V diagram for
  SBS~1129+577. The contour levels are at 5, 10, 15, 20, 25,
  30, 35, 40, 45, 49 mJy~beam$^{-1}$.}
\label{fig:lv}
\end{figure}

  The intensity-weighted velocity field (again at a resolution of
$\sim$40~arcsec) is shown in Fig.~\ref{fig:mom1}. The inner disc of
SBS~1129+576 appears to have a velocity field consistent with solid body
rotation, while the inner isovelocity contours of SBS~1129+577 are distorted.
We discuss this further below. Smooth velocity gradients can also be noted 
in the bridge and the one-armed spiral. The continuity in the velocity 
field is consistent with that expected from a prograde encounter between 
both galaxies. As expected in such an encounter, the rotational velocity 
gradient across SBS~1129+576 is in the same sense as that of the relative motion 
of SBS~1129+577. Similarly, the rotation velocity gradient across SBS~1129+577 
is in the same sense as that of the relative motion of SBS~1129+576. The 
morphology of the system (i.e. well defined tidal tails and a bridge) is also 
consistent with such an encounter geometry (see e.g. \cite{toomre}). 
The morphology and kinematics of the system also strongly resemble those 
of \ion{H}{i}~1225+01, which is also believed to be undergoing a prograde encounter 
(Chengalur, Giovanelli $\&$ Haynes 1995). Integrated spectra of
SBS~1129+576, SBS~1129+577, the bridge emission and the dwarf companion are
shown in Fig.~\ref{fig:spectra}, while the parameters derived from these
spectra are given in Table~\ref{tab:main}. The sum of the \ion{H}{i} flux from 
SBS~1129+576/SBS~1129+577 (i.e. 20.7~Jy~\kms) is in good agreement with
that measured by \cite{thuan99} (i.e. 21.6~$\pm$~0.8~Jy~\kms). We find
W$_{50}$ $\sim$85~\kms for SBS~1129+577 (which  makes the dominant 
contribution to the total \ion{H}{i} flux), again in good agreement with the 
value of 89~$\pm$~2~\kms given by \cite{thuan99}(1999). Position-velocity (P--V) 
diagrams along the major axis of SBS~1129+576 and SBS~1129+577 
are shown in Fig.~\ref{fig:lv}.

\begin{figure}
\includegraphics[width=8.5cm]{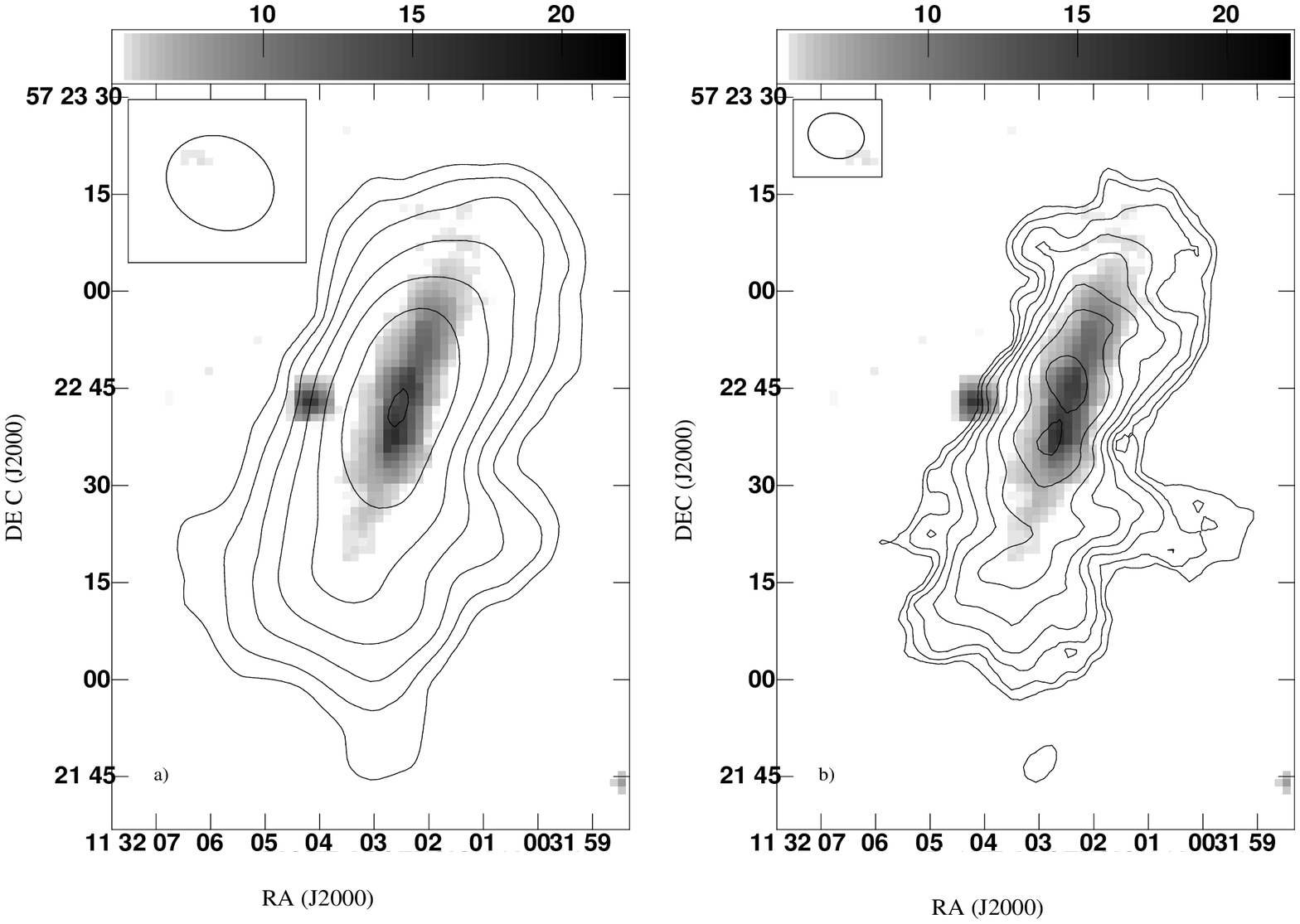}
\caption{\textbf{a)} Integrated \ion{H}{i} emission map (contours) of SBS~1129+576 at
   the resolution of 16~$\times$~14 arcsec$^{2}$ overlaid on the optical
   DSS~II $B$-band image (grey-scale in arbitrary units). The contour levels are at the column
   density of 3.4, 5.1, 7.7, 11.5, 17.3, 25.9, 38.9~$\times$~10$^{20}$
   atoms~cm$^{-2}$. \textbf{b)} Integrated \ion{H}{i} emission map of SBS~1129+576
   at the resolution of 8~$\times$~7~arcsec$^{2}$ in contours overlaid on
   the optical $B$-band image in (grey-scale in arbitrary units). 
The contour levels are at the
   column densities of 3.4, 5.1, 7.7, 11.5, 17.3, 25.9,
   38.9, 58.4~$\times$~10$^{20}$ atoms~cm$^{-2}$.}
\label{fig:highres}
\end{figure}

   As discussed above (see also Fig.~\ref{fig:mom0}), at a resolution of
$\sim$40~arcsec SBS~1129+576 shows at most a hint of a counter tail. 
Higher-resolution ($\sim$17~$\times$~14~arcsec$^{2}$ and 
8~$\times$~7~arcsec$^{2}$) maps of the \ion{H}{i} emission (overlaid on the
Digitized Sky Survey~II (DSS~II) $B$-band optical image) from this galaxy are shown in 
Fig.~\ref{fig:highres}. One can see that the \ion{H}{i} distribution
extends farther towards the south-east than towards the north-west. The
integrated spectrum of the galaxy (Fig.~\ref{fig:spectra}) also shows more
emission from gas at velocities corresponding to the southern half of the
galaxy. In Fig.~\ref{fig:highres}(a), one can also see that the contours 
near the southern tip of the optical disc extend towards the east. This
is suggestive of a warp. In Fig.~\ref{fig:highres}(b), one can see the
extension more clearly, and also that the \ion{H}{i} contours in the northern part
of the galaxy have a corresponding (but smaller) bend towards the opposite
direction, consistent with  an integral-shaped warp of the \ion{H}{i} disc. 
The gas in the northern end of the `warp' blend into the bridge
connecting the two galaxies.

\begin{figure}
\includegraphics[width=7.25cm,angle=270]{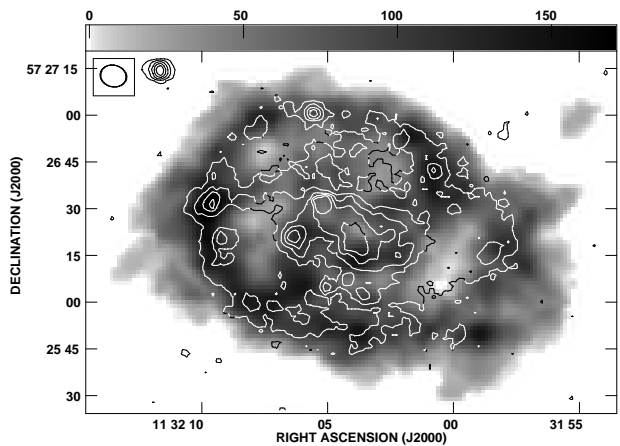}
\caption{The DSS~II optical blue band image (contours) overlaid on the
   integrated \ion{H}{i} emission map of SBS~1129+577 (grey-scale over the range of \ion{H}{i} 
column density from 0 to 3.6~$\times$~10$^{21}$ atoms~cm$^{-2}$) at a resolution of
   8~$\times$~7~arcsec$^{-2}$. The contours are logarithmically spaced and are
   in arbitrary units.}
\label{fig:highresB}
\end{figure}

 The integrated 8~$\times$~7-arcsec$^{2}$-resolution \ion{H}{i} map of
SBS~1129+577 (overlaid on the DSS~II $B$-band optical emission) 
is shown in Fig.~\ref{fig:highresB}. The optical emission is shown 
in contours, whereas the \ion{H}{i} emission is shown in grey-scale. The 
optical image shows a bar in the centre with a ring around it. 
The \ion{H}{i} emission is broadly similar; the main differences being 
that the ring is more continuous in \ion{H}{i} than in optical and that 
there is, in general, no clear correlation between the \ion{H}{i} column 
density variations and the intensity of the optical emission in the bar.

\begin{figure}
\includegraphics[width=7.25cm,angle=270]{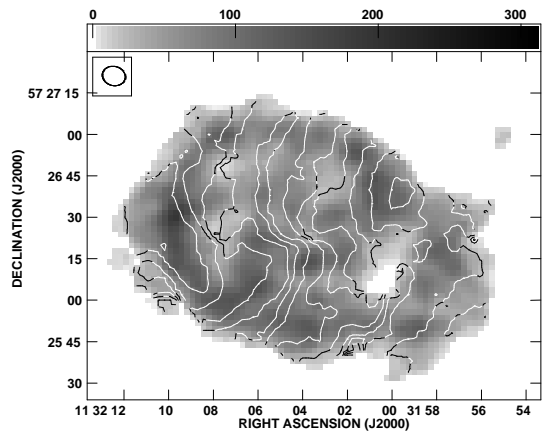}
\caption{ The velocity field of SBS~1129+577 at a resolution of
   8~$\times$~7~arcsec$^{-2}$ overlaid on the integrated \ion{H}{i} emission (grey-scale 
over range of \ion{H}{i} column density from 0 to 6.6~$\times$~10$^{21}$~atoms~cm$^{-2}$). 
The velocity contours are regularly spaced at an interval of 
6.6~\kms\ and range from 1528~\kms\ to 1607~\kms.}
\label{fig:highresmom1}
\end{figure}

The low-resolution velocity field of SBS~1129+577 (Fig.~\ref{fig:mom1}) 
suggests that the outer part of the gas disc is warped (compare e.g. 
with the velocity field of M51; \cite{rots90}). Further, as noted above, 
the central regions show distorted isovelocity contours. The kinematical 
major axis and kinematical minor axis are not perpendicular
to one another, as is typical for barred galaxies (e.g. NGC 5383; 
\cite{sancisi79}). The high-resolution (8~$\times$~7-arcsec$^{2}$) velocity 
field (Fig.~\ref{fig:highresmom1}) also shows that the isovelocity 
contours are not parallel to each other within the bar. This is suggestive 
of an oval distortion in the potential with the gas moving along 
elliptical orbits (cf. \cite{bosma81}). 

\par The velocity field of SBS~1129+576/577 system is very disturbed and
we could not fit a rotation curve to the emission from the individual
galaxies. We instead compute an indicative dynamical mass given by 
${\rm M}_{\rm ind} = 2.3 \times 10^{5} \times R_{\rm kpc} \times
V^2_{\rm kms^{-1}}$~M$_{\odot}$. The maximum velocity seen in the 
P--V diagram in Fig.~\ref{fig:lv} is 60~\kms. \cite{guseva03} (2003) gave
an axial ratio of 0.25 for this galaxy, which (assuming an intrinsic
axial ratio of 0.25) implies that it is being viewed very close to edge-on. 
We hence take the inclination-corrected  rotational 
velocity to be $\sim$60~\kms\ at radius of $\sim$8.3~kpc, giving
a dynamical mass of M$_{\rm dyn}$~$\sim$7.1$~\times$~10$^{9}$~M$_{\odot}$ 
for SBS~1129+576. Similarly, from Fig.~\ref{fig:lv} the maximum 
radial velocity for SBS~1129+577 is 65~\kms. For this galaxy, the
axial ratio listed in NASA/IPAC Extragalactic Database (NED) is $\sim$0.77, which corresponds to
an inclination of $\sim 42^o$ (again for an intrinsic axial 
ratio of 0.25). The maximum rotational velocity is hence
$\sim$88~\kms\ at radius of $\sim$12 kpc. This gives an 
indicative dynamical mass of M$_{\rm dyn}$~$\sim$2.1$~\times$~10$^{10}$~M$_{\odot}$.
 The slices used while deriving P--V diagrams were along kinematic major 
axis for both the galaxies.

    \section{DISCUSSION}
\label{sec:dis}

   In Table~\ref{tab:main}, we summarise the main observational and derived
parameters of SBS~1129+576 and its two companion galaxies. These two companion
galaxies are situated $\sim$27~kpc north (SBS~1129+577) and south-east 
(SDSS~J113227.68+572142.3) of the XMD galaxy, respectively. The total projected
size of this compact triplet is $\sim$43~kpc and the range of their radial 
velocities is $\sim$70~\kms. The oxygen abundance in SBS~1129+577 is 12~+~$\log$~(O/H)~=~7.94~$\pm$~0.08 
(based on Special Astrophysical Observatory (SAO) 6-m telescope observations; \cite{Pustilnik06} 2006). This value is
typical of BCGs with similar values of luminosity and suggests that it is an
old galaxy. In contrast, SBS~1129+576 has 12~+~$\log$~(O/H)~=~7.41~$\pm$~0.07 
(\cite{guseva03} 2003); there are only about a dozen or so known BCGs with 
metallicites comparable or lower than this. The third galaxy in the
group, SDSS~J113227.68+572142.3, has $g$ and $r$ magnitudes (as given in 
Sloan Digital Sky Survey (SDSS) Data Release 3 (DR3) data base; \cite{DR3}) of 16.95 and 16.81~mag, respectively. 
Using the formulae in \cite{smith02}, this translates to $B$~=~17.19~mag,
M$_{\rm B}$~=~--14.8~mag. Similarly, translating the SDSS $g$ and $r$ 
magnitudes for SBS~1129+576 and SBS~1129+577 to $B$ band, yields 
$B_{\rm tot}$ of 16.44 and 15.50~mag, respectively. Combining these 
with the \ion{H}{i} masses derived in this paper gives M$_{\rm HI}$/L$_{\rm B}$ 
of 2.1 and 3.8 (in solar units) for SBS~1129+576 and SBS~1129+577, 
respectively. Note that, in the value of M$_{\rm HI}$/L$_{\rm B}$
for SBS~1129+576 we included the mass of gas in the bridge, assuming that it
consists mainly of the gas pulled out of the less-massive galaxy.
These values of M$_{\rm HI}$/L$_{\rm B}$ are among the largest known for
starburst galaxies and confirm that both these galaxies are gas-rich. In
contrast, SDSS J113227.68+572142.3 is rather gas-poor, with
M$_{\rm HI}$/L$_{\rm B}$ = 0.08. The origin and evolution of such a system 
of quite different types of dwarf galaxies is probably worth a 
separate deeper study.

\subsection{Star formation in SBS~1129+576, SBS~1129+577 and SDSS~J113227.68+572142.3}

   As mentioned in Section~\ref{sec:intro}, \cite{guseva03} (2003) presented detailed
models of the star formation history of SBS~1129+576. Most of the current
star formation is concentrated in two bright knots of emission (called
regions 'a' and 'b' in \cite{guseva03} 2003). These two bright knots can also be
seen in the broad-band optical image in Fig.~\ref{fig:highres}, where it can
also be seen that they coincide with peaks in the \ion{H}{i} column density
distribution. From the measured metallicity, the H$\alpha$ equivalent widths (EWs),
and also assuming a standard Salpeter initial mass function (IMF), \cite{guseva03} (2003) estimate that
these star-forming regions are very young, with ages $\lesssim$10~Myr. Like
in many other BCD galaxies, the current star formation activity
in SBS~1129+576 appears to be occurring only above a critical \ion{H}{i} threshold
density. The observed peak \ion{H}{i} column densities corresponding to \ion{H}{ii} regions
'a' and 'b' is $\sim$5.9~$\times$~10$^{21}$ and
$\sim$6.7~$\times$~10$^{21}$~atoms~cm$^{-2}$, respectively, at an angular
resolution of 8~arcsec which corresponds to a linear resolution of 1.02~kpc
at the assumed distance of 26.3~Mpc. These observed peak \ion{H}{i} column
densities correspond to deprojected values of the total gas density
(i.e. including a 10 per cent contribution from He, we note that this is a slight
overestimate of the He abundance in metal-poor galaxies) of 
$\sim$1.6~$\times$~10$^{21}$ and $\sim$1.8~$\times$~10$^{21}$~atoms~cm$^{-2}$,
similar to the values seen for other BCDs and dwarf irregular galaxies 
(e.g. \cite{skillman87}; \cite{taylor94}). 

\begin{table*}
\centering
\caption{\label{tab:main} Main observational and derived parameters of the 
studied dwarf galaxies}
\begin{tabular}{lrrr} \\ \hline 
Parameter               & SBS~1129+576     & SBS~1129+577     & SDSS~J113227+572142  \\ \hline
$\alpha\,$(J2000)       & 11$^{h}$32$^{m}$02$^{s}$.54 & 11$^{h}$32$^{m}$04$^{s}$.00  & 11$^{h}$32$^{m}$27$^{s}$.68\\
$\delta\,$(J2000) & +57\degr22\arcmin45\arcsec.7 & +57\degr26\arcmin20\arcsec.0 & +57\degr21\arcmin42\arcsec.3 \\
$\int f(v)dv$ (Jy~\kms) & 3.9$^{a}$              & 16.7             & 0.07           \\ 
$V_{hel}^{\,b}$ (\kms)  & 1506~$\pm$~3.3     &  1575~$\pm$~3.3    & 1540:    \\
W$_{\rm 50}$ (\kms)     &   67~$\pm$~3.3     &     85~$\pm$~3.3   &  18:           \\
m$_{B}^{\;\;c}$         &   16.44          &     15.50        & 17.19          \\
M$_{B}^{\;\;d}$         &   $-$15.65       &     $-$16.59     & $-$14.8        \\
$M_{\rm HI}$ (10$^8$M$_{\odot}$)&  6.3$^a$        &     27.0         &  0.11          \\
$M_{\rm HI}$/L$_{\rm B}$$^e$   &  2.1$^a$         &     3.8          &  0.08          \\
12~+~$\log$~(O/H)          & 7.41~$\pm$~0.07    & 7.94~$\pm$~0.08    & --             \\
\hline 
\\[-0.35cm]
\multicolumn{4}{l}{$^a$ \ion{H}{i} flux/mass of the bridge is included; $^b$ Heliocentric \ion{H}{i} velocity;} \\
\multicolumn{4}{l}{$^c$ Translated from SDSS $g$ and $r$ magnitudes;} \\
\multicolumn{4}{l}{$^d$ Absolute blue magnitudes corrected for the MW extinction and D=26.3 Mpc;} \\
\multicolumn{4}{l}{$^e$ In solar units.} \\
\end{tabular}
\end{table*}

    The distribution of star-forming regions in SBS~1129+577 is more
complicated. A long-slit spectroscopy of this galaxy was done by
\cite{Pustilnik06} (2006), using the SAO 6-m telescope. The slit was positioned
at a position angle of 90\degr\ and passing through the brightest \ion{H}{ii} region
A1 (J113206.5+572620; $\sim$20~arcsec to the east of the centre of the main
body seen in continuum). The peak \ion{H}{i} column density at this position is 
2.8~$\times$~10$^{21}$ atoms~cm$^{-2}$ at an angular resolution of 8~arcsec. 
Besides this region, the star formation is also seen to take place in three 
more regions along the slit, namely A2, A3 and A4. A2 is $\sim$20~arcsec 
to the east of A1, while A3 is $\sim$35~arcsec to the west of A1. The \ion{H}{i} 
column densities are
3.1~$\times$~10$^{21}$ and 2.4~$\times$~10$^{21}$~atoms~cm$^{-2}$
at the positions of A2 and A3, respectively. A4, which is rather faint \ion{H}{ii}
region near the centre of the galaxy, has an \ion{H}{i} peak with column density of
3.2~$\times$~10$^{21}$~atoms~cm$^{-2}$. We note that there are other brighter
knots visible in the continuum optical image which also coincide with peaks
in \ion{H}{i}. For example, there is a bright optical condensation
(J11329.9+572630) in the ring which coincides with the highest peak
(4.1~$\times$~10$^{21}$~atoms~cm$^{-2}$) in the \ion{H}{i} column density map. These
too may be star-forming regions, but the confirmation by H$\alpha$-imaging or
spectroscopy is required. For the \ion{H}{ii} regions identified above via the
long-slit spectroscopy, based on their measured EW(H$\beta$), we
estimate ages of 4--8~Myr for the case of an instantaneous starburst
model with the standard Salpeter IMF (\cite{Leitherer} 1999). The inclination-corrected 
total gas column densities for regions A1, A2, A3, and A4
are 2.3~$\times$~10$^{21}$, 2.6~$\times$~10$^{21}$, 2.0~$\times$~10$^{21}$,
and 2.6~$\times$~10$^{21}$, respectively, all of which are similar
to the canonical threshold density.

     As mentioned above, SDSS~J113227.68+572142.3 is gas-poor, with
$M_{\rm HI}$/L$_{\rm B}$~=~0.08. It does not appear to be strongly interacting with
the other two galaxies and, in particular, there is no evidence for the
exchange of matter between it and the two brighter galaxies of the group.
The SAO 6-m telescope spectrum (unpublished) for this object as well as its
SDSS DR3 spectrum show an H$\alpha$ emission line with EW~=~28~\AA, a faint
[\ion{O}{iii}]$\lambda$5007 line and Balmer absorption lines in blue. According to
the Starburst-99 models of \cite{Leitherer} (1999), for an instantaneous starburst
with the metallicity $z$~=~0.0004 and a standard Salpeter IMF, all of these features 
are consistent with a star formation episode with an age of
$\sim$20~Myr. This age is larger but comparable with that of the
starbursts in SBS~1129+576 and SBS~1129+577. For SDSS~J113227.68+572142.3, 
the peak \ion{H}{i} column density (at a resolution of $\sim$8~arcsec or 1~kpc) 
is $\sim$3.9~$\times$~10$^{20}$~atoms~cm$^{-2}$, corresponding to a deprojected
total gas column density of $\sim$4.0~$\times$~10$^{20}$~atoms~cm$^{-2}$. This
is a factor of $\sim$2 below the nominal star formation threshold. 
We note that \cite{begum} also found that star formation in very faint
dwarf galaxies does occur at gas column densitites approximately two times lower
than the nominal threshold density. The size of the galaxy is 
comparable to our beam size, even at the highest resolution (i.e. 
$\sim$8~arcsec). It is thus very likely that our observed peak 
column density is a severe underestimate of the true column density.
It is possible that the star
formation in SDSS J113227.68+572142.3 was triggered by the accretion of a
small amount of gas onto a gas-poor (dwarf elliptical ?) progenitor.
Deeper \ion{H}{i} observations may help to confirm or rule out this hypothesis.

\subsection{Comparison with other XMD galaxies}

   The SBS~1129+576/577 system is morphologically very similar to \ion{H}{i}~1225+01
(\cite{salzer92} 1992, \cite{chengalur95}). Both systems consist of two
components with a bridge between the two. In both systems, the southern
component is almost edge-on, whereas the northern component is more face-on. The
\ion{H}{i} finger at the south-western edge of SBS~1129+576 is a countertail similar to that in
the south-western component of \ion{H}{i}~1225+01. The north-east clump of \ion{H}{i}~1225+01 has a bar
as SBS~1129+577 does have. Like in \ion{H}{i}~1225+01, we assume that the \ion{H}{i} 
in the bridge originates from the smaller galaxy in the pair. SBS~1129+576, 
like the south-western component in \ion{H}{i}~1225+01, is on a prograde
orbit. The \ion{H}{i} masses of the less-massive components in both systems are
similar (where we assume a distance of 15~Mpc to \ion{H}{i}~1225+01). The ratio of
the masses of the two components is, however, different, with a value of
$\sim$1:1.9 for \ion{H}{i}~1225+01 and $\sim$1:4.3 for SBS~1129+576/577. In computing
these ratios, we assume that \ion{H}{i} in the bridges originally belonged to less-massive components. 
The major difference between the systems SBS~1129+576/577
and \ion{H}{i}~1225+01 is that no optical counterpart has been found in south-west clump of
\ion{H}{i}~1225+01 to a limiting surface  brightness of 27.0
$V$-mag~arcsec$^{-2}$ (\cite{salzer91}).

    It is interesting to compare the  environments of these systems.
\cite{salzer92} (1992) found that there is no dwarf galaxy within $\sim$1~Mpc of
\ion{H}{i}~1225+01, while the nearest luminous galaxy is at $\sim$1.2~Mpc from it (for
an adopted distance of 15~Mpc). We re-examined the environment of \ion{H}{i}~1225+01,
using the data available in NED. We found that only one dwarf VCC~1208 (which
is 124~arcmin north of \ion{H}{i}~1225+01 and has $\delta V$~=~7~\kms) lies within the
$\sim$1~Mpc limit given by \cite{salzer92} (1992). This galaxy lies at a projected
distance of 540~kpc (for  D~=~15~Mpc). 
The region around SBS~1129+576 was
surveyed as a part of SDSS. Since there are no
galaxies with 1300~\kms\ $<$ V$_{\rm hel}$ $<$ 1700~\kms\ within 30~arcmin
(corresponding to the projected distance of 230~kpc), it seems to be an
isolated group of dwarfs.
There are several galaxies at the projected distances of 150--500 kpc with
the velocities in the range $\sim$1700--1800~\kms\ (including a rather
luminous SB(s)c galaxy NGC~3683, with $M_{\rm B}$ of $-$19.6~mag), but their
radial velocities imply that all of them are more than
$\sim$2~Mpc away from the SBS~1129+576/577 system. There are four dwarf galaxies
with radial velocities from 1450 to 1565~\kms\ at the projected distances
of $\sim$350--400 kpc. The nearest luminous galaxy with close radial
velocity is NGC~3619
($M_{\rm B} = -$19.5~mag) at $\sim$0.8~Mpc. We conclude that both systems are
at a distance of $\sim$1~Mpc from the nearest luminous galaxies and at
several hundred kpc from the nearest dwarf galaxy.

\par SBS~0335-052 (\cite{pustilnik01} 2001a) is another interesting object to
compare to the SBS~1129+576/577 system. Both the components, SBS~0335$-$052 E and W, at
a projected distance of 22~kpc (\cite{pustilnik01} 2001a) are candidate young
galaxies (e.g. \cite{izotov97}; \cite{papaderos98}; Pustilnik, Pramskij \& Kniazev 2004; \cite{izotov05}).
The \ion{H}{i} envelope surrounding these components is similar to the one
seen in SBS~1129+576/577 system. Its total gas mass is comparable to that of
SBS~1129+577 (i.e. $\sim$10$^{9}$~M$_{\odot}$). The morphology and kinematics
of SBS~0335-052 in \ion{H}{i} and at other wavelengths suggest that the most
plausible interpretation of this system is a merger. The progenitors are
either extremely unevolved very gas-rich objects or protogalaxies. Their
mutual tidal interaction has triggered the current
bursts of star formation. The peak measured \ion{H}{i} column densities are
1.0~$\times$~10$^{21}$ and 0.7~$\times$~10$^{21}$~atoms~cm$^{-2}$ in the 
west and east components, respectively, at a linear
resolution of 5.4~$\times$~3.9~kpc. This system, however,  seems to be
not very isolated, lying at the outskirts of the loose group LGG 103.
SBS~0335-052 is at a projected distance of $\sim$150 kpc, and a velocity
difference of $\sim$120~\kms\ from the luminous spiral NGC~1376.

  Similarly, the current starburst in the low-metallicity galaxy HS~0822+3542
appears to be triggered by interaction with its companion SAO~0822+3545
(\cite{pustilnik03}; \cite{chengalur06}; but 
see \cite{corbin05} for an alternative interpretation). For I~Zw~18,
an XMD system with the second-lowest known metallicity, there
is an extended \ion{H}{i} distribution with four distinct components
(\cite{vanzee98}), only two of which have optical counterparts.

    Interpreting the observed configuration of SBS~1129+576 as the first
close encounter (as is for the case of \ion{H}{i}~1225+01), we would expect that the
significant exchange of matter has not  yet taken place. This is
consistent with the large difference of metallicities (O/H differ by a factor
of $\sim$3, or 0.5~dex). As noted by \cite{pustilnik04}, in some of the
probable XMD mergers, which are at more advanced stages, the differences in
O/H in different components are significantly smaller. This is consistent
with the gas exchange leading to a more homogeneous composition of 
ISM in the merging components towards the late stages of
merger. In particular, for HS~0122+0743 (\cite{pramskij03}; Pustilnik et
al., in preparation), in which optical components are in contact,
O/H differs by less than 0.1~dex. On the other hand, for SBS~0335--052 
E and W, which seem to have their first close encounter the 
situation appears to be more complicated. While the galaxies are not
expected to have exchanged much gas, their O/H differs by only 0.17 dex.
This may be the result of the metal-enrichment by the first star formation episodes 
in both components leading to a typical value of Z~$\sim$1/30~Z$_{\odot}$ 
(\cite{kunth86}).
Summarising the comparison of SBS~1129+576 with several other XMD BCGs, we
find that the tidal-interaction-triggered star formation is a common theme
for all of them.
 
\section{Summary}
\label{sec:summary}

We summarise our conclusions about this \ion{H}{i} study of the metal-poor galaxy
SBS~1129+576 and its immediate surroundings as follows:

\begin{enumerate}

\item SBS~1129+576 is undergoing a strong interaction with the nearby
   dwarf galaxy SBS~1129+577. We assume that \ion{H}{i} has been pulled out 
from it to form a bridge
   between the two galaxies. The mass of \ion{H}{i} in the bridge is 2.1~$\times$
   10$^{8}$~M$_{\odot}$, roughly half the \ion{H}{i} mass remaining in the disc 
   of SBS~1129+576. The
   observed morphology of the system suggests that it is in a prograde orbit.
   The system is probably on its first encounter and no significant exchange
   of material has taken place yet. This is consistent with the very different
   ISM metallicities of the two systems. The enhanced star formation in these
   gas-rich galaxies ($M_{\rm HI}$/L$_{\rm B}$ of 2.1 and 3.8) has very likely been
   triggered by mutual tidal interaction.

\item  The two \ion{H}{i} peaks in SBS 1129+576 are at face-on surface densities of
       $\sim$1.5~$\times$~10$^{21}$ and $\sim$1.7~$\times$~10$^{21}$~atoms~cm$^{-2}$. 
These coincide with the positions of the two
       brightest \ion{H}{ii} regions, respectively. The face-on \ion{H}{i} column density in
       \ion{H}{ii} regions identified in SBS~1129+577 using the slit spectroscopy is
       above $\sim$1.8~$\times$~10~$^{21}$~atoms~cm$^{-2}$. This is similar in
       value to the threshold density for star formation seen in other BCGs.

\item  SBS~1129+577 has a central bar with a ring surrounding it, while the
       outer gas has been distorted into a strong one-armed spiral pattern.
       These are also likely to be consequences of the tidal interaction.

\item   The velocity field of SBS~1129+577 shows that the disc is warped in
	the outer region, while in the inner region the gas orbits appear to
	be non-circular, probably as a response to the potential of the
	central bar. There is a hint of an integral-shaped warp in the 
inner parts of disc of SBS~1129+576.

\item   The system has a third dwarf galaxy with \ion{H}{i} mass of
	1.1$\times$10$^{7}$ M$_{\odot}$. It has an $M_{\rm HI}$/L$_{\rm B}$ value
	of 0.08, that is, much lower than those for its more-massive companions.
	While the optical spectrum of this galaxy indicates a recent
	star-formation episode 
	($\sim$20 Myr, comparable to that for starburst ages in SBS~1129+576
	and SBS~1129+577), its low $M_{\rm HI}$/L$_{\rm B}$ suggests that it has had
	a very different evolutionary history. It is further unusual in having 
ongoing star formation, despite having a peak observed gas column density lower 
than the nominal threshold for star formation. We note, however, the true column 
density may be considerably higher than that derived from our comparatively low 
poor resolution maps. 

\item   The system SBS~1129+576/577 is similar in many aspects to two
	 other well-known XMD systems. These are \ion{H}{i}~1225+01 and 
         SBS~0335--052 E and W, where the current star burst appears to
         have been triggered by tidal interaction.

\end{enumerate}

{\it Acknowledgment.}
SAP acknowledges the partial support from the Russian federal program
'Non-stationary phenomena in astronomy'.
The authors acknowledge the spectral and photometric information from
the SDSS data base, used for this study.
The Sloan Digital Sky Survey (SDSS) is a joint project of the University of
Chicago, Fermilab, the Institute for Advanced Study, the Japan Participation
Group, the Johns Hopkins University, the Max-Planck-Institute for Astronomy
(MPIA), the Max-Planck-Institute for Astrophysics (MPA), New Mexico State
University, Princeton University, the United States Naval Observatory, and
the University of Washington. Apache Point Observatory, site of the SDSS
telescopes, is operated by the Astrophysical Research Consortium (ARC). 
 We thank the GMRT staff for making these observations possible. The
GMRT is run by the National Centre for Radio Astrophysics of the
Tata Institute of Fundamental Research.

\label{lastpage}

\end{document}